\documentclass[epj]{webofc}
\usepackage[varg]{txfonts}   
\usepackage{lineno}

\let\oldenumerate\enumerate
\renewcommand{\enumerate}{
  \oldenumerate
  \setlength{\itemsep}{1pt}
  \setlength{\parskip}{0pt}
  \setlength{\parsep}{0pt}
}
\let\olditemize\itemize
\renewcommand{\itemize}{
  \olditemize
  \setlength{\itemsep}{1pt}
  \setlength{\parskip}{0pt}
  \setlength{\parsep}{0pt}
}

\woctitle{6th Rome International Conference on AstroParticle Physics}
\begin{document}
%
%
\title{Rare Events searches with Cherenkov Telescopes}

\author{
  Michele Doro\inst{1}\fnsep\thanks{\email{michele.doro@pd.infn.it}}
}
\institute{
University and INFN Padova, via Marzolo 8, I-35131 Padova (Italy)
}

\abstract{%
Ground-based Imaging Cherenkov Telescope Arrays observe the Cherenkov
radiation emitted in extended atmospheric showers generated by cosmic
gamma rays in the TeV regime. The rate of these events is normally
overwhelmed by 2--3 orders of magnitude more abundant cosmic
rays induced showers. A large fraction of these ``background'' events
is vetoed at the on-line trigger level, but a substantial fraction
still goes through data acquisition system and is saved for the off-line
reconstruction. What kind of information those events carry,
normally rejected in the analysis? Is there the possibility that an
exotic signature is hidden in those data? In the contribution, some
science cases, and the problems related to the event reconstruction
for the current and future generation of these telescopes will be
discussed.   
}
\maketitle
\section{Introduction}
Gamma-ray Astronomy is the branch of science that observes the cosmic
radiation beyond the keV. Below some tens of GeV, such observation is
done mostly through pair-production instruments (e.g. the Fermi-LAT
instrument\footnote{\url{fermi.gsfc.nasa.gov/}}) or Compton-scattering
instruments, mounted on satellites. Above few tens of GeV and below
several tens of TeV, observations are mostly done with Imaging
Atmospheric Cherenkov Telescope Arrays 
(IACTA) that observe indirectly gamma rays through the Cherenkov light
produced by atmospheric particle shower initiated in the high Earth
atmosphere by cosmic gamma rays. Despite this technique has only 3
decades now, it has already reached a mature
stage~\cite{Lorenz:2012nw,Hillas:2013txa}, with about 150 sources
detected, and a world-wide installation soon to be deployed, under the
name of CTA (Cherenkov Telescope Array~\cite{CTA:2010}). 

There are currently three major installations of IACTA: H.E.S.S.,
MAGIC and VERITAS\footnote{H.E.S.S.: \url{www.mpi-hd.mpg.de/hfm/HESS/},
  MAGIC: \url{wwwmagic. mppmu.mpg.de}, VERITAS: \url{veritas.sao.arizona.edu}} that are under operations for
about a decade now. These instruments perform stereoscopic
observations of the same event with multiple telescopes: the Cherenkov
radiation from the atmospheric shower, generates, on the
cameras of the telescopes, an ellipse-like shape, whose image
treatment allow inferring the direction and energy of the corresponding
primary
cosmic gamma ray. To image an event, IACTA cameras are constituted by
more than a thousand pixel each (the individual pixel is typically a
photomultiplier tube of typically $0.1$~deg aperture). In such instruments, there are several layers
of triggers and selection of events, some acting online, some
offline. The first levels need to exclude the noise events caused by
the Light of the Night Sky, due to  starlight, zodiacal light and airglow. This is done online. A
rate of about 200 Hz of events passing this selection is typically 
stored on disk\footnote{For the ten years of operation of MAGIC, considering
an average datataking of 5~h per night, this rate corresponds to about
12 GEvents saved on disk}. However, most of these events do not
correspond to gamma rays, but instead are comprised of atmospheric shower
events initiated by cosmic rays (mostly protons, with traces of
heavier nuclei). The hadronic background at this stage outnumbers the
gamma-ray events by more than a factor of hundred. Later on during the
data reconstruction, these hadronic events are rejected by further
image cleaning and selection. However, not all background can be
rejected, specially at the lowest energies, where the images are more
dim. 

In this contribution, we briefly discuss the possibility that some of the background
events can have actually a different origin, in some cases even hiding
signatures of exotic and fundamental physics. We argue that one can
develop special reconstruction and analysis treatment to extract these
events. We are motivated to discuss this issue by gathering together
different phenomena, for two reasons: from one side, the search of
hidden signals in the background data of IACTA share similarities
(special image cleaning, special data selection, whole data sample
access, blind signal searches), and from the other side, it could be
timely to consider fast selection filter for the CTA
instrument. The reason is that, while current IACTA can manage to save
data
on disk because the space occupation is limited (about 1~TB of data/day
for, e.g., MAGIC), for CTA the situation will be more dramatic, with expected
100 TB data/day or even more. In order to reduce the occupancy, CTA is
planning to preselect and delete some information on the events. If
this will not be done efficiently, CTA will risk to throw away
possible extremely interesting events in its data haystack. 

A
search for such needles in the haystack would require several dedicated steps in the
reconstruction and analysis: 
\begin{enumerate}
\item A dedicated Monte Carlo. All events of IACTA are determined by
  comparing the image in the multipixel camera with the corresponding
  Monte Carlo simulation. For gamma rays and hadrons, this is done using the
  Corsika code. For peculiar events, one should additionally develop a
  code for the interaction of the cosmic particle with the
  atmosphere. It is clear that in some cases, when an exotic particle
  is under scrutiny, such Monte Carlo will be not only complex to
  develop, but will rely on theoretical ansatz; 
\item A dedicated image cleaning. The standard image cleaning
  (although different techniques were proposed in the past)
  relies on the extrapolation of the event image by ``cleaning out'' those
  pixels whose signal is very likely caused by the Light of the
  Night Sky. The procedure is optimized for ellipse-like shapes (like
  those coming from gamma rays) through the so-called Hillas parameterization~\cite{Hillas:1985a}. Some rare events could have instead
  very peculiar images (small bright spots, multiple images, etc, -- see below). A
  dedicated procedure should thus be prepared; 
\item A dedicated parameterization of the event and extraction of
  primary information (direction, energy); 
\item A dedicated high-level analysis.
\end{enumerate}

It is clear that the finding of one event will very likely not be
sufficient to infer a detection. All rare events should happen with
sufficient statistics to be visible above an unresolvable background. 

\section{Rare Events in the Background sample}
The first class of rare phenomena that will be discussed is composed of
events that have passed the first on-telescope trigger criteria,
have been rejected by the standard analysis, and are stored on the
disks. Of these events, some could have a 
classic nature, some could belong to more exotic explanations.

\subsection{Heavy Nuclei}
In 2007, H.E.S.S. reported the measurement of the spectrum of cosmic
iron nuclei from 13 TeV to 200 TeV~\cite{Aharonian:2007zja} with five
spectral points. Their data nicely overlap previous measurements taken
with balloon experiments. Events from iron nuclei are two orders of
magnitude less frequent than proton-induced shower, and for this
reasons are harder to detect. However, the signatures in IACTA
from heavy-nuclei-initiated showers have different features
than those of proton or gamma rays. The nucleus is charged
and proceeding with relativistic speed. Therefore, a small but
intense burst of Cherenkov radiation is \emph{directly} produced by
the nucleus itself in the high atmosphere. As soon as it travels down, the
nucleus has interactions with the denser atmospheres initiating an hadronic shower, rather similar to that of the protons. Therefore,
in the camera, an iron event is composed of two spots: a bright spot
toward the center of the camera (high in the atmosphere) arriving
earlier, followed later in time by the classical ellipsoidal shape of
the atmospheric showers and aligned with the main shower axis. The analysis is not straightforward, but
proven possible. Besides H.E.S.S., no other IACTA has tested this
method.

One could ask whether other heavy ions can be seen in the cosmic ray
spectrum, whose abundances per element are measured at lower energies
with balloons (see, e.g.~\cite{Hu:2009sn}). Particles like CNO or Si
are not only rarer because of lower fluxes, but also would provide less photon yield
(that goes as Z$^2$). However, specially with future generation of
telescopes like CTA, with better sensitivity and larger energy range,
such searches will be possible.

\subsection{Tau-Neutrino searches}
Several classes of astronomical targets including massive black holes
at the center of active galaxies or gamma-ray bursts, are expected to
produce significant radiation of neutrinos. Irrespective of the family
of neutrino at the production place, for extragalactic distances, the
mixing foresees that the neutrino families at the earth should
arrive in equal fraction, and thus that cosmic tau-neutrinos should be
observable at Earth. These have not yet been discovered in cosmic neutrino detectors,
however, they may be observable with IACTA through a phenomenon
called Earth-skimming taus~\cite{Asaoka20137,Fargion:2007gi}. Shortly, if a tau-neutrino crosses the
right amount of ground (the Earth crust, or water), of the order of
few tens of km, tau-leptons can be generated through deep inelastic
scattering processes like
$\nu_\tau + N \rightarrow W^+ \rightarrow X+\tau^-$ If
the tau-lepton later on emerges from the medium, it 
creates an atmospheric shower. Suppose now that a telescope is located
at the right distance from the exit point of the tau-lepton, it could
detect the emerging atmospheric shower. From such directions, a shower
could be not explained by other mechanisms. Searches like this were performed by MAGIC
looking at the right direction toward the Canarian sea, reporting for now only results
on the feasibility of the technique, but still no detection~\cite{Gaug:2007fpa}. The
expectation on the flux are extremely low: the diffuse neutrino flux
can provide few events per decade. However, in case of strong or flaring
astrophysical sources, the neutrino flux could be enhanced, thus providing still
 dim, but detectable signals. When one then compares the sensitivity of
e.g. CTA compared to other instruments like Auger or IceCube, one can
see that for ``low-energy'' PeV neutrinos, the CTA sensitivity could be
larger than the others, thus providing sufficient ground for a careful
search~\cite{Gora:2016mmy}. MAGIC developed the selection criteria for these events,
showing that tau-neutrino induced events are in principle observable in the
data. 

\subsection{Magnetic Monopoles}
Magnetic monopoles were predicted back in 1930 by Dirac to explain the
electric quantization. Later on during the century, it was
found that magnetic monopoles appear naturally in Grand Unification
models~\cite{Hooft:1974,Polyakov:1974}. In particular, some theories predict
that they are formed during the QCD phase transition in the early
Universe, and, being stable, they could still be present in the actual
Universe. When a
magnetic monopole crosses the Earth atmosphere, it will produce a huge
number of Cherenkov photons, about 4700 times than those produced by a
gamma ray~\cite{Tompkins:1965}. In addition, the Cherenkov photons
from magnetic monopoles 
will be produced throughout
the full length of the atmosphere, and not from a limited path as
when originated by atmospheric showers. This event would be observable by
an IACTA as extremely bright spots or short lines, and not like
ellipse-like shapes. The search for magnetic monopoles events has already been accomplished by H.E.S.S.~\cite{Spengler:2011}
and the expectation for CTA were discussed
in~\cite{Doro:2012xx}. However, other instruments like Auger or
IceCube seems to
have higher sensitivity~\cite{Fujii:2015hxp,Aartsen:2015exf}. One
should also mention that IACTA would be sensitive only to
ultrarelativistic magnetic monopoles, while other instruments have
wider capabilities~\cite{Aartsen:2014awd}.

\subsection{Antiquark Matter}
In order to explain the matter-antimatter density inequality in the
present Universe, some theories predict that during baryogenesis,
the antimatter content was confined into very high dense states of
quark plasma by the formation and subsequent collapse of domain walls
in the existing quark-gluon 
plasma~\cite{Zhitnitsky:2002qa,Oaknin:2003uv}. Such aggregation would
be composed of a huge number of antiquarks (or quarks), in the order of
$10^{25}-10^{35}$, and have survived until present times in the
intergalactic medium. These aggregation are called ``quark nuggets''
and share similarities with the strangelets~\cite{Farhi:1984}. They can be
considered as viable dark matter candidate, at least comprising a
fraction of the total density. The quark nuggets would be
dressed with leptons to be globally neutral. In
several works of K.~Lawson, and specially~\cite{Lawson:2015mpa}, the
direct and indirect detection techniques for quark nuggets are
described. In particular, the quark nuggets are 
expected to emit charged particles and high-energy radiation when crossing the
Earth atmosphere, thus initiating an extended atmospheric shower. The
main difference with respect to standard cosmic showers would stem
from the fact that the nugget will not decay in the atmosphere, and that
its velocity is much lower than that of cosmic rays, typically of the
order of the galactic velocities. The passage will then be seen as a
``stripe'' on the camera of the telescope, developing slowly from one
side of the camera to the other, considering the nugget velocity, and
increasing in brightness toward the ground, where the nuggets
interactions with the denser atmosphere would increase.

No dedicated search for these exotic states has been performed with
IACTs so far. However, the search would share similarities with the
case of magnetic monopoles, as discussed in~\cite{Aartsen:2014awd}.

\section{Rare Events in the Field of View}
Not only one can have peculiar events in the background data haystack, with
specific signatures in duration, time evolution, shape, etc., as
described in the previous section, but additional rare events could
occur serendipitously within the field of view, passing undetected,
unless a specific analysis is developed. It is clear that a steady
source or very brilliant flaring source in the field of view is recognized
through the standard analysis. Here we are discussing examples of very
brief events, lasting seconds or less, that would not appear when
integrating over larger time windows. 

\subsection{Primordial Black Hole Evaporation}
There are several mechanisms that allow the creation of primordial black holes
(PBHs) in the Early Universe, besides those of astrophysical origin. Depending
on the Universe average density at a given time after the Big
Bang, these PBHs could have a specific mass. However, the range of
possible masses is large: from the Planck mass to
$10^5\;\mathrm{M}_\odot$~\cite{Carr:2009jm}. As the time passes, a PBH
increase its temperature and radiate energy, toward the final phase
when Hawking radiation is emitted, and
the BH evaporates. The life expectancy of a BH can be computed and depends
solely on the BH mass. As the evaporation time approaches, the BH
radiates more on more. This means that, at present times, we could be
seeing the evaporation of all the PBHs of a \emph{given} mass.
A description of the lightcurve and gamma-ray
spectrum of emission from an exploding PBH can be found
in~\cite{MacGibbon:2015mya}. Shortly, the gamma-ray emission would be
stable for most of the time, while an exponential increase in the last minutes to seconds
to the evaporation is predicted.

PBH evaporation could be therefore appear as short bursts of emission
randomly in the FOV of an IACTA regular observation. Bright and steady
sources in the FOV are in principle easily seen in these
data. However, in this case the emission would be more subtle to find
and its observation would require a dedicated analysis: the emission could be dim, and
specially short in time, and therefore washed out by integration over
large duration. The PBH search should be performed over the whole
data sample of an IACTA. This requires some non-standard data
handling. The Whipple gamma-ray telescope pioneered this
search~\cite{Linton:2006yu}, however, 
much better sensitivity can be expected with CTA.

\subsection{Fast Radio Bursts}
Fast Radio Bursts are very short ($1-10$ ms) bursts of radiation
discovered in archival radio data few years
ago~\cite{Lorimer:2007qn}. Besides their short duration, the main
characteristics is that the radio emission shows a large wavelength
dispersion, which hints to extragalactic origin ($z\sim1$). They
could be originated out of neutron stars or magnetars formation or
merger events~\cite{Murase:2016sqo}. These peculiar extreme events
recently raised attention in the astrophysics community, however,
their true nature is still to be clarified. When computing the
intensity, it is possible that these events are accompanied by the
emission of gamma-ray radiation at TeV energies. Very similarly to the
PBH case discussed above, they would therefore appear as very short
and intense spots in IACTA skymaps. The light
curve should be different from that of PBH, so they could be
discriminated. No result is published yet from IACTA in search of
these targets.


\section{Discussion and Conclusions}


In this contribution, we have briefly discussed few possibilities to
search for rare events among the data gathered by ground-based
gamma-ray detectors of the IACTA class. These rare events share some
features: they would probably go undetected by standard
reconstruction and analysis techniques, thus they would require
dedicated simulations, data selection, image treatments and so on. We
grouped these events into two classes: events that would be mostly
tagged as background in the IACTA standard reconstruction, and events
that would appear serendipitously in the field of view of the instrument,
and could go unnoticed because of short duration or faintness.

Past and current instruments have performed searches and published
results around some of these topics, some instead are not investigated
yet. These projects share complexity in terms of data handling and
often suffer from incomplete theoretical mapping. However, in most
cases, these investigations would not require 
allocation of instrumental time. They would mostly imply a careful
treatment the large dataset of archival data gathered by the current
instruments, which is now comprised of about a decade of data.

Because the current instruments generate a large but not huge amount
of data per night, basically all data that triggered the telescopes
are safely stored on disks, including a large fraction of background
(cosmic ray) data which is normally partly unused in the
advanced steps of the analysis. However, the future large installation of CTA will
produce a very large amount of data per night, which would demand an
effort to reduce consistently the full information, e.g. by excluding
some pixels from the image event, or reducing the amount of
background-tagged events stored on disk. For gamma-ray searches, this
is an optimal solution, but for the search of rare events proposed
here, this could be a killer factor. It is therefore envisaged to
develop robust and fast routines that 
could tag interesting not-standard events and save them for
further analysis. It is clear that such routines should be developed
well on time before CTA starts operation, which is expected soon after
2020. 

\begin{acknowledgement}
I want to thank A.~Morselli for the invitation at the conference,
M.~Martinez for having pushed me to investigate these topics some
years ago and collect some ideas, A. De Angelis, D. Gora and J. Rico
and the anonymous referee for comments on the text.  
\end{acknowledgement}

 \bibliography{doro_bib.bib}
 
\end{document}